\newif\ifpdf
\ifx\pdfoutput\undefined
    \pdffalse
\else
    \pdfoutput=1
    \pdffalse
\fi
\documentclass[12pt]{article}
\ifpdf
    \usepackage[pdftex]{graphicx}
    \usepackage[pdftex]{hyperref}
\else
    \usepackage{graphicx}
    \usepackage{hyperref}
\fi

\usepackage{scicite}

\usepackage{times}

\newenvironment{sciabstract}{%
\begin{quote} \bf}
{\end{quote}}

\newcounter{lastnote}
\newenvironment{scilastnote}{%
\setcounter{lastnote}{\value{enumiv}}%
\addtocounter{lastnote}{+1}%
\begin{list}%
{\arabic{lastnote}.}
{\setlength{\leftmargin}{.22in}}
{\setlength{\labelsep}{.5em}}}
{\end{list}}

\title{`Deconfined' quantum critical points}

\author
{T. Senthil,$^{1\ast}$ Ashvin Vishwanath,$^{1}$ Leon Balents,$^{2}$   \\
Subir Sachdev,$^{3}$ Matthew P. A. Fisher$^{4}$ \\
\\
\normalsize{$^{1}$Department of Physics, Massachusetts Institute
of Technology, Cambridge MA 02139}\\
\normalsize{$^{2}$Department of Physics, University of California,
Santa Barbara, CA 93106-4030}\\
\normalsize{$^{3}$Department of Physics, Yale University, P.O. Box
208120, New Haven, CT 06520-8120}\\
\normalsize{$^{4}$Kavli Institute for Theoretical Physics, University of
California, Santa Barbara, CA 93106-4030}\\
\\
\normalsize{$^\ast$To whom correspondence should be addressed;
E-mail:  senthil@mit.edu.} }

\date{September 22, 2003}

\begin{document}
\ifpdf
    \DeclareGraphicsExtensions{.pdf}
\fi


\baselineskip24pt


\maketitle


\begin{sciabstract}
The theory of second order phase transitions is one of the
foundations of modern statistical mechanics and condensed matter
theory. A central concept is the observable `order parameter',
whose non-zero average value characterizes one or more phases and
usually breaks a symmetry of the Hamiltonian. At large distances
and long times, fluctuations of the order parameter(s) are
described by a continuum field theory, and these dominate the
physics near such phase transitions. In this paper we show that
near second order {\em quantum } phase transitions, subtle quantum
interference effects can invalidate this paradigm. We present a
theory of quantum critical points in a variety of experimentally
relevant two-dimensional antiferromagnets. The critical points
separate phases characterized by conventional `confining' order
parameters. Nevertheless, the critical theory contains a new
emergent gauge field, and `deconfined' degrees of freedom
associated with fractionalization of the order parameters. We
suggest that this new paradigm for quantum criticality may be the
key to resolving a number of experimental puzzles in correlated
electron systems.
\end{sciabstract}

\section*{Introduction and motivation}

Much recent research has focused attention on the behavior of matter
near zero temperature `quantum' phase transitions that are seen in
several strongly correlated many particle systems\cite{subirbook}.
Indeed, a currently popular view ascribes many properties of
interesting correlated materials to the competition between
qualitatively distinct ground states and the associated phase
transitions.  Examples of such materials include the cuprate
high-$T_c$ superconductors, and the rare-earth intermetallic compounds
(known as the heavy fermion materials).

The traditional guiding principle behind the modern theory of
critical phenomena is the association of the critical
singularities with fluctuations of an `order parameter' which
encapsulates the difference between the two phases on either side
of the critical point (a simple example is the average magnetic
moment which distinguishes ferromagnetic iron at room temperature,
from its high temperature paramagnetic state). This idea,
developed by Ginzburg and Landau \cite{landau}, has been eminently
successful in describing a wide variety of phase transition
phenomena. It culminated in the sophisticated renormalization
group theory of Wilson \cite{wilson}, which gave a general
prescription for understanding the critical singularities. Such an
approach has been adapted to examine quantum critical phenomena as
well, and provides the generally accepted framework for
theoretical descriptions of quantum transitions.

The purpose of this paper is to demonstrate and study specific
examples of quantum phase transitions which do not fit into this
Ginzburg-Landau-Wilson (GLW) paradigm \cite{oned}. We will show
that in a number of different quantum transitions, the natural
field theoretic description of the critical singularities is not
in terms of the order parameter field(s) that describe the bulk
phases, but in terms of new  degrees of freedom specific to the
critical point. In the examples studied in this paper, there is an
emergent gauge field which mediates interactions between emergent
particles that carry fractions of the quantum numbers of the
underlying degrees of freedom. These `fractional' particles are
not present ({\em i.e.\/} are confined) at low energies on either
side of the transition, but appear naturally at the transition
point. Laughlin has previously argued for fractionalization at
quantum critical points on phenomenological grounds \cite{rbl}.

The specific situations studied in this paper are most
conveniently viewed as describing phase transitions in two
dimensional quantum magnetism, although other applications are
also possible\cite{Crtny}.  Consider a system of spin $S =
1/2$ moments $\vec S_r$ on the sites, $r$, of a two dimensional
square lattice with the Hamiltonian
\begin{equation}
H = J\sum_{\langle rr'\rangle}\vec S_r\cdot \vec S_{r'} + ....
\label{ham}
\end{equation}
The ellipses represent other short ranged interactions that may be
tuned to drive various zero temperature phase transitions.  We
assume $J > 0$, {\em i.e} predominantly antiferromagnetic interactions.

\begin{figure}
\centerline{\includegraphics[width=4in]{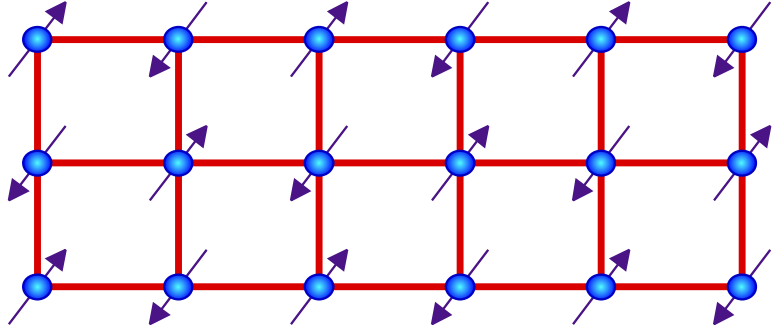}} \caption{The
magnetic N\'{e}el ground state of the Hamiltonian
(\protect\ref{ham}) on the square lattice. The spins, $\vec S_r$,
fluctuate quantum mechanically in the ground state, but they have
a non-zero average magnetic moment which is oriented along the
directions shown.} \label{neel}
\end{figure}

\begin{figure}
\centerline{\includegraphics[width=4.8in]{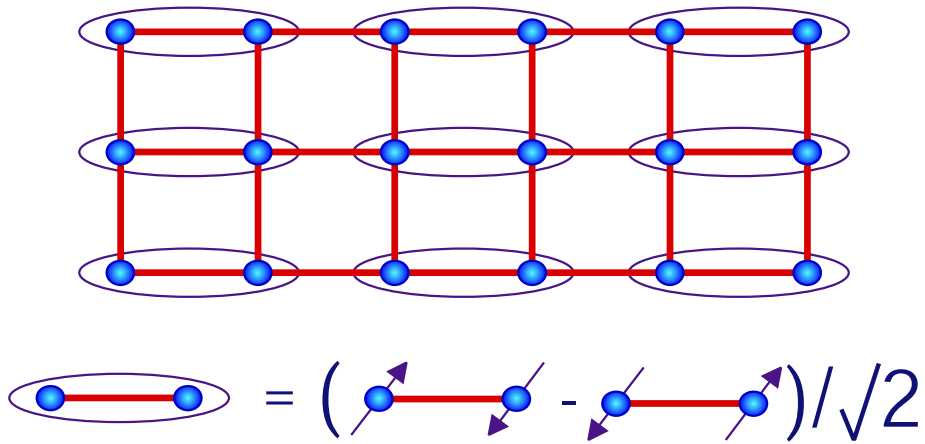}} \caption{A
valence bond solid (VBS) quantum paramagnet. The spins are paired
in singlet valence bonds, which resonate among the many different
ways the spins can be paired up. The valence bonds `crystallize',
so that the pattern of bonds shown has a larger weight in the
ground state wavefunction than its symmetry-related partners
(obtained by 90$^\circ$ rotations of the above states about a
site). This ground state is therefore four-fold degenerate.}
\label{bond}
\end{figure}

Considerable progress has been made over the last fifteen years in
elucidating the nature of the various possible ground states of such a
Hamiltonian. The N\'{e}el state has long range magnetic order (see
Fig.~\ref{neel}) and has been observed in a variety of insulators
including the prominent parent compound of the cuprates:
La$_2$CuO$_4$.
Apart from such magnetic states, it is now recognized that models
in the class of $H$ can exhibit a variety of quantum paramagnetic
ground states. In such states, quantum fluctuations prevent the
spins from developing magnetic long range order. Such paramagnetic
states can be broadly subdivided into two groups. First, there are
states that can be described as `valence bond solids' (VBS)
--- a simple example is shown in Fig~\ref{bond}.
In such states pairs
of nearby spins form a singlet, resulting in an ordered pattern of
`valence bonds'. Typically, such VBS states have an energy gap to
spin-carrying excitations.  Furthermore, for spin-$1/2$ systems on
a square lattice, such states also necessarily break lattice
translational symmetry. A second class of more exotic paramagnetic
states are also possible\cite{rvb} in principle: in these states,
the quantum-mechanical resonance between different valence bond
configurations is strong enough to disrupt the VBS, and we obtain
a resonating-valence-bond `liquid'. The resulting state has been
argued to possess excitations with fractional spin and interesting
topological structure \cite{krs,ReSaSpN,Wen,sf}. In this paper we
will {\em not} consider such exotic paramagnetic states.  Rather,
our focus will be on the nature of the phase transition between
the ordered magnet and a VBS. We will also restrict our discussion
to the simplest kinds of ordered antiferromagnets - those with
collinear order where the order parameter is a single vector (the
N\'{e}el vector).

Both the magnetic N\'{e}el state, and the VBS are states of broken
symmetry. The former breaks spin rotation symmetry, and the latter
that of lattice translations. The order parameters associated with
these two different broken symmetries are very different.  A
simple Landau-like description of the competition between these
two kinds of orders generically predicts either a first-order
transition, or an intermediate region of coexistence where both
orders are simultaneously present. A direct second order
transition between these two broken symmetry phases requires
fine-tuning to a `multicritical' point.  Our central thesis is
that for a variety of physically relevant {\em quantum} systems,
such canonical predictions of Landau's theory are incorrect. For
$H$, we will show that a generic second order transition is
possible between the very different kinds of broken symmetry in
the N\'{e}el and VBS phases. Our critical theory for this
transition is, however, unusual, and is {\em not} naturally
described in terms of the order parameter fields of either phase.
Although we will not explore this case further here, a picture
related to the one developed here applies also to
transitions between fractionalized spin liquid and VBS states\cite{long}, and
to transitions between different VBS states\cite{vbspap} in the quantum dimer
model \cite{rk,sondhi}.

\section*{Field theory and topology of quantum antiferromagnets}

In the N\'{e}el phase or close to it, the fluctuations of the
N\'{e}el order parameter are captured correctly by the well-known
O(3) nonlinear sigma model field theory\cite{hald88,chn} with the
following action in spacetime (we have promoted the lattice
co-ordinate $r=(x,y)$ to a continuum spatial co-ordinate, and $\tau$ is
imaginary time):
\begin{equation}
\mathcal{S}_n = \frac{1}{2g} \int d\tau \int d^2 r
\left[\frac{1}{c^2}
  \left(\frac{\partial \hat n}{\partial \tau}\right)^2 +
  \left(\nabla_r
    \hat n \right)^2 \right] + i S \sum_r (-1)^r \mathcal{A}_r .
    \label{nls}
\end{equation}
Here $\hat n \propto (-1)^r \vec S_r$ is a unit three component
vector that represents the N\'eel order parameter (the factor
$(-1)^r$ is $+1$ on one checkerboard  sublattice, and $-1$ on the
other). The second term is the quantum mechanical Berry phase of
all the $S=1/2$ spins: $\mathcal{A}_r$ is the area enclosed by the
path mapped by the time evolution of $\hat{n}_r$ on a unit sphere
in spin space. These Berry phases play an unimportant role in the
low energy properties of the N\'{e}el phase \cite{chn}, but are
crucial in correctly describing the quantum paramagnetic phase
\cite{ReSaSuN}. We will show here that they also modify the
quantum critical point between these phases, so that the exponents
are distinct from those of the GLW theory without Berry phases
studied earlier \cite{chn,csy}.

\begin{figure}
\centerline{\includegraphics[width=5in]{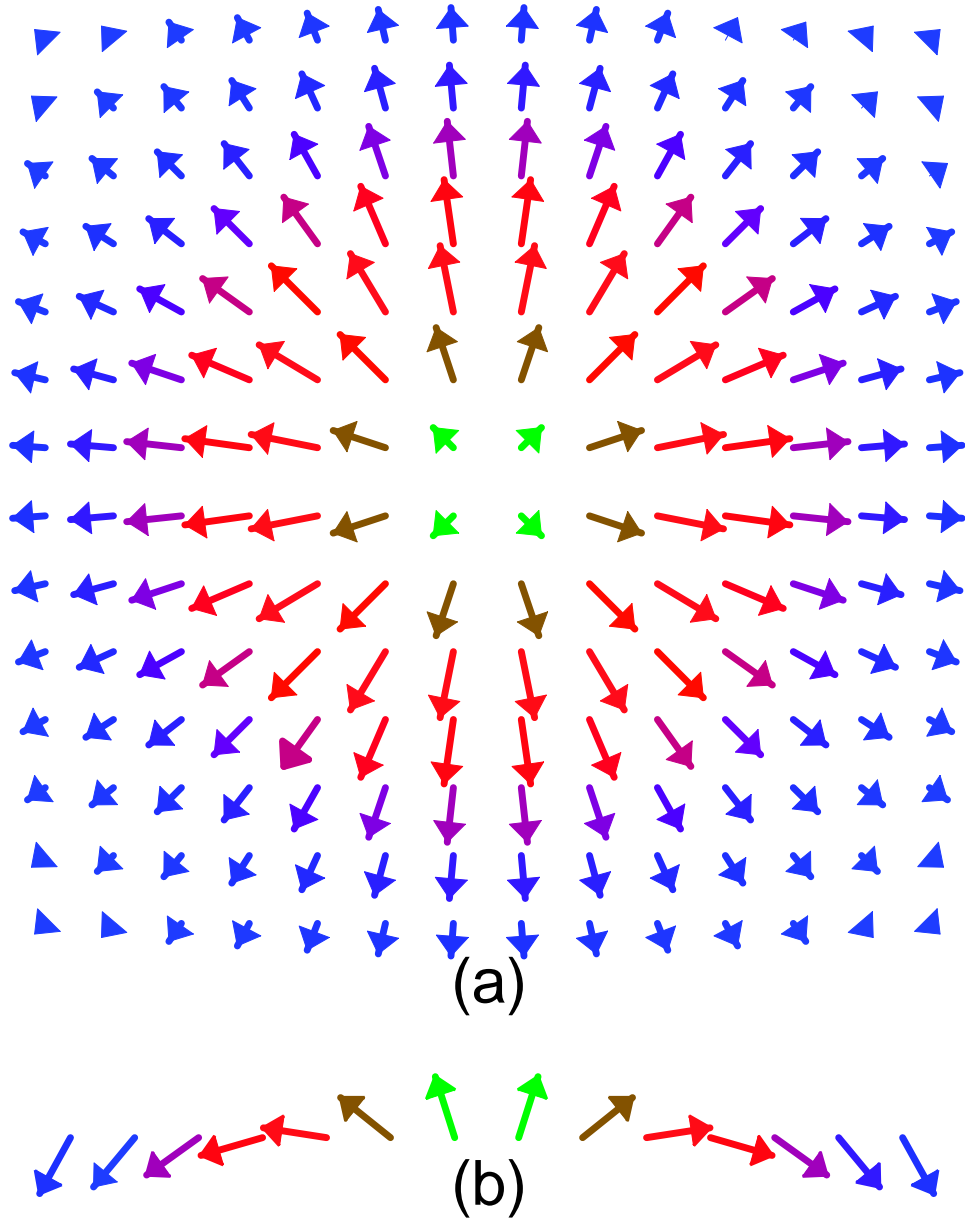}} \caption{A skyrmion
configuration of the field $\hat{n}(r)$. In (a) we show the vector
$(n^x,n^y)$ at different points in the $xy$ plane; note that
$\hat{n} = (-1)^r \vec S_r$, and so the underlying spins have a
rapid sublattice oscillation which is not shown. In (b) we show
the vector $(n^x, n^z)$ along a section of (a) on the $x$ axis.
Along any other section of (a), a picture similar to (b) pertains,
as the former is invariant under rotations about the $z$ axis. The
skyrmion above has $\hat{n} (r=0) = (0,0,1)$ and $\hat{n}(|r|
\rightarrow \infty) = (0,0,-1)$.} \label{skyr}
\end{figure}

\begin{figure}
\centerline{\includegraphics[width=4in]{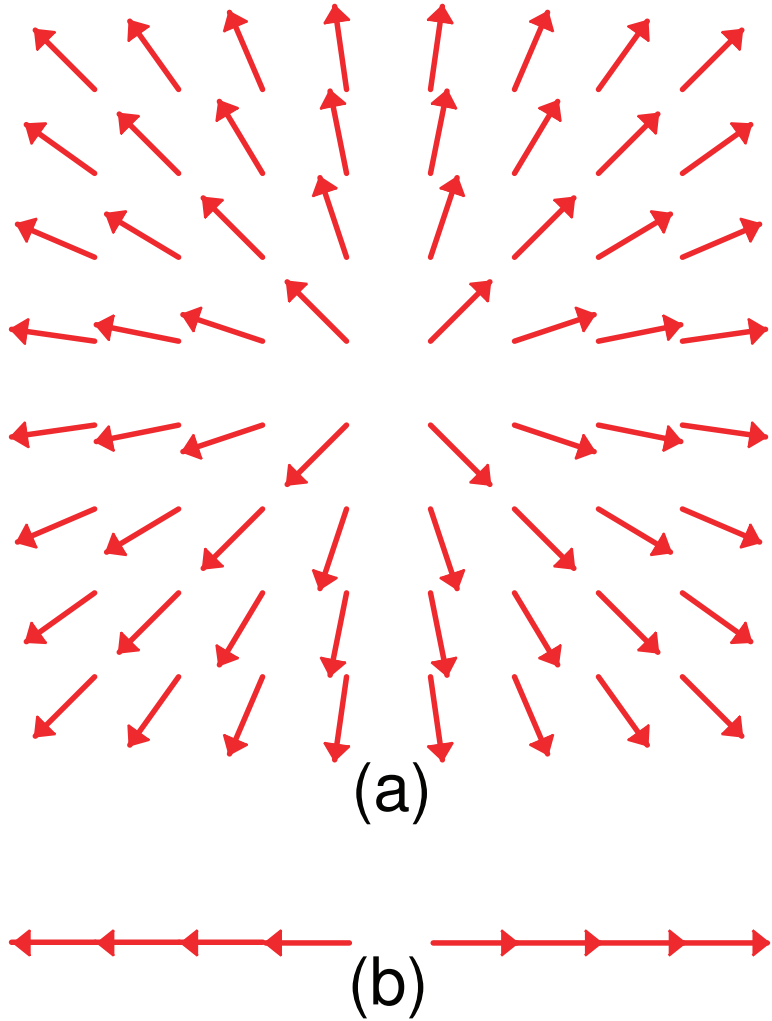}} \caption{A
monopole event, taken to occur at the origin of spacetime. An
equal-time slice of spacetime at the tunnelling time is
represented following the conventions of Fig~\ref{skyr}. So (a)
contains the vector $(n^x, n^y)$; the spin configuration is
radially symmetric, and consequently a similar picture is obtained
along any other plane passing through the origin. Similarly, (b)
is the representation of $(n^x, n^z)$ along the $x$ axis, and a
similar picture is obtained along any line in spacetime passing
through the origin. The monopole above has $\hat{n} (r) = r/|r|$.}
\label{hedge}
\end{figure}

To describe the Berry phases, first note that in two spatial
dimensions,
smooth configurations of the N\'{e}el vector admit topological
textures known as skyrmions (see Fig~\ref{skyr}).
The total {\em
skyrmion number} associated with a configuration defines an
integer topological quantum number $Q$:
\begin{equation}
\label{poynt} Q = \frac{1}{4\pi} \int d^2 r \, \hat n \cdot
\partial_x \hat{n} \times \partial_y \hat{n},
\end{equation}
The sum over $r$ in (\ref{nls}) vanishes\cite{hald88,subirbook}
for all spin time histories with {\em smooth} equal-time
configurations, even if they contain skyrmions. For such smooth
configurations, the total skyrmion number $Q$ is independent of
time. However, the original microscopic model is defined on a
lattice, and processes where $Q$ changes by some integer amount
are allowed. Specifically, such a $Q$ changing event corresponds
to a monopole (or `hedgehog') singularity of the N\'{e}el field
$\hat n(r, \tau)$ in space-time (see Fig~\ref{hedge}).
Haldane\cite{hald88} showed that the sum over $r$ in (\ref{nls})
is non-vanishing in the presence of such monopole events. Precise
calculation\cite{hald88} gives a total Berry phase associated with
each such $Q$ changing process which oscillates rapidly on four
sublattices of the dual lattice. This leads to destructive
interference
which effectively suppresses all monopole events unless they are
quadrupled\cite{hald88,ReSaSuN} ({\em i.e} they change $Q$ by
four).

The sigma model field theory augmented by these Berry phase terms
is, in principle, powerful enough to correctly describe the
quantum paramagnet.  Summing over the various monopole tunnelling
events shows that in the paramagnetic phase the presence of the
Berry phases leads to VBS order\cite{ReSaSuN}. Thus
$\mathcal{S}_n$ contains within it the ingredients describing both
the ordered phases of $H$.
However a description of the transition between these phases has so
far proved elusive, and will be provided here.

Our analysis of this critical point is aided by writing the N\'eel
field $\hat n$ in the so-called CP$^1$ parametrization:
\begin{equation}
\label{cp1} \hat n =  z^\dagger \vec \sigma z^{\vphantom\dagger} ,
\label{nz}
\end{equation}
with $\vec{\sigma}$ a vector of Pauli matrices. Here $z = z(r ,
\tau) = (z_1, z_2)$ is a two-component complex spinor of unit
magnitude which transforms under the spin-$1/2$ representation of
the SU(2) group of spin rotations. The $z_{1,2}$ are the
fractionalized ``spinon" fields. To understand the monopoles in
this representation, let us recall that the CP$^1$ representation
has a U(1) gauge redundancy. Specifically the {\em local} phase
rotation
\begin{equation}
z \rightarrow e^{i\gamma(r, \tau)}z , \label{zgauge}
\end{equation}
leaves $\hat n$ invariant, and hence is a gauge degree of freedom.
Thus the spinon fields are coupled to a U(1) gauge field, $a_\mu$
(the spacetime index $\mu = (r, \tau)$). As is well-known, the
magnetic flux of $a_\mu$ is the topological charge density of
$\hat n$ appearing in the integrand of (\ref{poynt}).
Specifically, configurations where the $a_\mu$ flux is $2\pi$
correspond to a full skyrmion (in the ordered N\'{e}el phase).
Thus the monopole events described above are space-time `magnetic'
monopoles (instantons) of $a_\mu$ at which $2\pi$ gauge flux can
either disappear or be created. That such instanton events are
allowed, means that the $a_\mu$ gauge field is to be regarded as
`{\em compact}'.

We now state our key result for the critical theory between the
N\'{e}el and VBS phases. As we will demonstrate below, the Berry
phase-induced quadrupling of monopole events renders monopoles
{\em irrelevant} at the quantum critical point. So in the critical
regime (but {\em not} away from it in the paramagnetic phase), we
may neglect the compactness of $a_\mu$, and write down the
simplest continuum theory of spinons interacting with a
non-compact U(1) gauge field with action $\mathcal{S}_z = \int d^2
r d \tau \mathcal{L}_z$, and
\begin{equation}
\label{sz} \mathcal{L}_{z} =  \sum_{a = 1}^N |\left(\partial_{\mu}
- ia_{\mu}\right) z_{a}|^2 + s |z|^2 + u\left(|z|^2 \right)^2  +
\kappa\left(\epsilon_{\mu\nu\kappa}\partial_{\nu}a_{\kappa}\right)^2,
\end{equation}
where $N=2$ is the number of $z$ components (later we will
consider the case of general $N$), we have softened the length
constraint on the spinons with $|z|^2 \equiv \sum_{a=1}^N |z_a |^2
$ allowed to fluctuate, and the value of $s$ is to be tuned so
that $\mathcal{L}_z$ is at its scale-invariant critical point.
Note that the irrelevance of monopole tunneling events at the
critical fixed point implies that the total gauge flux $\int d^2 r
(\partial_x a_y - \partial_y a_x)$, or equivalently the skyrmion
number $Q$, is asymptotically conserved. This emergent global
topological conservation law provides precise meaning to the
notion of deconfinement. It is important to note that the critical
theory described by $\mathcal{L}_z$ \cite{hlm} is distinct from
the GLW critical theory of the O(3) non-linear sigma model
obtained from (\ref{nls}) by dropping the Berry phases and tuning
$g$ to a critical value \cite{mv}. In particular, the latter model
has a non-zero rate of monopole tunneling events at the
transition, so that the global skyrmion number $Q$ is not
conserved.


A justification for the origin of $\mathcal{L}_z$ is provided in
the remainder of this paper. We will begin in the following
section by considering antiferromagnets with an `easy plane'
anisotropy, so that the spins prefer to lie in the $xy$ plane.
Subsequent sections generalize the arguments to fully isotropic
antiferromagnets.

\section*{Duality transformations with easy plane anisotropy}

For the easy-plane case, duality maps and an explicit derivation
of a dual form of $\mathcal{L}_z$ are already available in the
literature \cite{Crtny,sp}. Moreover, an easy plane $S=1/2$ model
with two- and four-particle ring exchanges has recently been
studied numerically\cite{sandvik}, and a direct transition between
N\'{e}el and VBS phases was found. Here we will obtain the theory
for this transition \cite{Crtny,sp} using simple physical
arguments which enable generalization to the isotropic case.

The easy plane anisotropy reduces the continuous SU(2) spin
rotational invariance to the U(1) subgroup of rotations about the
$z$-axis of spin.  An additional important discrete symmetry is
time reversal, under which
\begin{equation}
\vec S_r \rightarrow -\vec{S}_r.
\end{equation}
This may be combined with a rotation in the $xy$ plane which
restores the sign of $S_{x,y}$ to simply change the sign of $S^z$
alone, comprising a $Z_2$ symmetry. With these symmetries,
(\ref{nls}) allows an additional term  $u_{ep} \int d\tau d^2 r
(n^z)^2$ , with $u_{ep} > 0$.

Let us first think classically about this easy plane model. The
classical (N\'eel) ground state simply consists of letting $\hat n$ be
independent of position, and lying entirely in the spin $xy$ plane.
Topological defects in this ground
state will play an important role. With the easy plane anisotropy,
these are vortices in the complex field $n^+ = n_x + in_y$. More
precisely, on going around a large loop containing a vortex the
phase of $n^+$ winds around by $2\pi m$ with $m$ an integer.

\begin{figure}
\centerline{\includegraphics[width=4in]{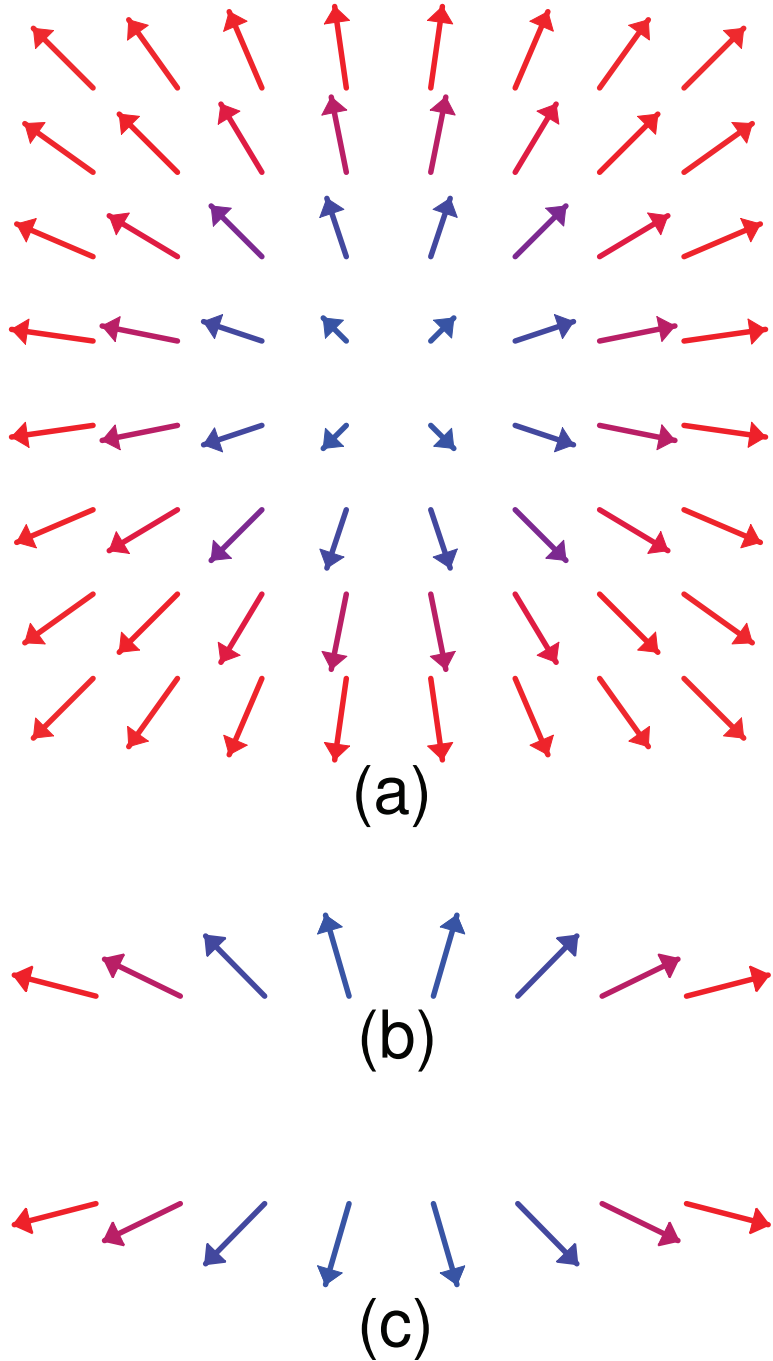}} \caption{The `meron'
vortices in the easy plane case. There are two such vortices,
$\psi_{1,2}$, and $\psi_1$ is represented in (a) and (b), while
$\psi_2$ is represented by (a) and (c), following the conventions
of Fig~\protect\ref{skyr}. The $\psi_1$ meron above has $\hat{n}
(r=0) = (0,0,1)$ and $\hat{n} ( |r| \rightarrow \infty ) =
(x,y,0)/|r|$; the $\psi_2$ meron has $\hat{n} (r=0) = (0,0,-1)$
and the same limit as $|r| \rightarrow \infty$. Each meron above
is `half' the skyrmion in Fig~\protect\ref{skyr}: this is evident
from a comparison of (b) and (c) above with
Fig~\protect\ref{skyr}b. Similarly, one can observe that a
composite of $\psi_1$ and $\psi_2^\ast$ makes one skyrmion.}
\label{mer}
\end{figure}

What is the nature of the core of these vortices? In the core, the
XY order will be suppressed and the $\hat n$ vector will point
along the $\pm \hat z$ direction. Physically,
this corresponds to a non-zero staggered magnetization of
the $z$-component of the spin in the core region.  Thus, at the
classical level there are two kinds of vortices -- often called `merons'
in this context -- depending on the
direction of the $\hat n$ vector at the core (see Fig~\ref{mer}).
Either kind of vortex {\em breaks} the Ising-like $n^z \rightarrow
-n^z$ symmetry at the core. For future convenience, let us denote
by $\psi_1$ ($\psi_2$) the quantum field that destroys a  vortex whose core
points in the up (down) direction.

Clearly, this breaking of the Ising symmetry is an artifact of the
classical limit - once quantum effects are included, the two
broken symmetry cores will be able to tunnel into each other and
there will be no true broken Ising symmetry in the core.  This
tunnelling is often called an `instanton' process that connects
two classically degenerate states.

Surprisingly, such an instanton event is physically the easy plane
avatar of the space-time monopole described above for the fully
isotropic model. This may be seen pictorially. Each classical
vortex of Fig~\ref{mer} really represents half of the skyrmion
configuration of Fig~\ref{skyr}. Now imagine a $\psi_2$ meron at
time $\tau \rightarrow -\infty$ with a spatial configuration as in
Fig~\ref{mer}a,c, and the $\psi_1$ meron as in Fig~\ref{mer}a,b at
time $\tau \rightarrow \infty$. These two configurations cannot be
smoothly connected, and there must be a singularity in the
$\hat{n}$ configuration, which we place at the origin of
spacetime. A little imagination now shows that the resulting
configuration of $\hat{n}$ can be smoothly distorted into the
radially symmetric monopole event of Fig~\ref{hedge} (indeed, the
union of Figs~\ref{mer}b,c placed as shown is easily seen to be
similar to Fig~\ref{hedge}a). Thus, the tunnelling process between
the two merons is equivalent to creating a full skyrmion. This is
precisely the monopole event of Fig~\ref{hedge}. Hence, as
pictorially reinforced in Figs.~\ref{skyr},\ref{mer}, a skyrmion
may be regarded as a composite of an ``up'' meron and a ``down''
antimeron, and the skyrmion number is hence the difference in the
numbers of up and down merons.

The picture so far has not accounted for the Berry phases terms.
The interference effect discussed above for isotropic antiferromagnets
applies here too, leading to an effective cancellation of instanton
tunnelling events between single $\psi_1$ and $\psi_2$ merons.  The
only effective tunnelling are those in which {\em four\/} $\psi_1$
merons come together and collectively flip their core spins to produce
four $\psi_2$ merons, or vice versa.

A different perspective on the $\psi_{1,2}$ meron vortices is
provided by the CP$^1$ representation. Ordering in the $xy$ plane
of spin-space requires condensing the spinons,
\begin{equation}
| \langle z_1\rangle |= |\langle z_2\rangle | \neq 0 ,
\end{equation}
so that $n^+ = z^*_1 z_2^{\vphantom*}$ is ordered {\em and} there
is no average value of $n^z = |z_1|^2 - |z_2|^2$.  Now, clearly, a
full $2\pi$ vortex in $n^+$ can be achieved by either having a
$2\pi$ vortex in $z_1$ and not in $z_2$, or a $2\pi$ antivortex in
$z_2$ and no vorticity in $z_1$. In the first choice, the
amplitude of the $z_1$ condensate will be suppressed at the core,
but $\langle z_2\rangle$ will be unaffected. Consequently $n^z =
|z_1|^2 - |z_2|^2$ will be non-zero and negative in the core, as
in the $\psi_2$ meron. The other choice also leads to non-zero
$n^z$ which will now be positive, as in the $\psi_1$ meron.
Clearly, we may identify the $\psi_2$ ($\psi_1$) meron vortices with
$2\pi$ vortices (anti-vortices) in the spinon fields $z_1$ ($z_2$).
Note that in terms of the spinons, paramagnetic phases correspond
to situations in which neither spinon field is condensed.

The above considerations, and the general principles of boson
duality in three spacetime dimensions \cite{dh} determine
the form of the dual action, $\mathcal{S}_{\rm dual}=\int \! d\tau d^2
r\, \mathcal{L}_{\rm dual}$ for $\psi_{1,2}$ \cite{Crtny,sp}:
\begin{eqnarray}
\label{crtny} \mathcal{L}_{\rm dual}& = & \sum_{a =
1,2}|\left(\partial_{\mu} - iA_{\mu}\right)\psi_{a}|^2 + r_d
|\psi|^2 + u_d \left(|\psi|^2 \right)^2 \nonumber \\
&& \hspace{-0.5in} + v_d |\psi_1|^2 |\psi_2|^2 + \kappa_d
\left(\epsilon_{\mu\nu\kappa}\partial_{\nu} A_{\kappa}\right)^2
 - \lambda [\left(\psi^*_1 \psi^{\vphantom{*}}_2 \right)^4 +
 \left(\psi^*_2 \psi^{\vphantom{*}}_1 \right)^4 ],
\end{eqnarray}
where $|\psi|^2 \equiv |\psi_1|^2 + |\psi_2|^2$.

The correctness of this form may be argued as follows: First, from the
usual boson-vortex duality transformation\cite{dh}, the dual $\psi_{1,2}$
vortex fields must be minimally coupled to a dual {\em
non-compact} U(1) gauge field $A_\mu$. Note that this dual gauge
invariance is {\em not} related to (\ref{zgauge}), but is a
consequence of the conservation of the total $S^z$: the `magnetic'
flux $\epsilon_{\mu\nu\kappa}
\partial_\nu A_\kappa$ is the conserved $S^z$ current \cite{dh}.
Second, under the $Z_2$ $T$-reversal symmetry
the two vortices get interchanged, {\em i.e} $\psi_1
\leftrightarrow \psi_2$. The dual action must therefore be
invariant under interchange of the $1$ and $2$ labels. Finally, if
monopole events were to be disallowed by hand, the total skyrmion
number -- {\em i.e.\/} the difference in number of up and down
meron vortices -- would be conserved.
This would imply a global U(1) symmetry (not to be confused with
the U(1) spin symmetry about the $z$ axis) under which
\begin{equation}
\label{dglobu1} \psi_1  \rightarrow  \psi_1 \exp(i\alpha)~~;~~
\psi_2  \rightarrow  \psi_2 \exp(-i\alpha) ,
\end{equation}
where $\alpha$ is a constant. However, monopole events destroy the
conservation of skyrmion number, and hence this dual global U(1)
symmetry. But, as the monopoles are effectively quadrupled by
cancellations from Berry phases, {\em skyrmion number is still
conserved modulo 4}. Thus the symmetry (\ref{dglobu1}) must be
broken down to the discrete cyclic group of four elements, $Z_4$.

The dual Lagrangian in (\ref{crtny}) is the simplest one that is
consistent with all these requirements. In particular, we note
that in the absence of the $\lambda$ term the dual global U(1)
transformation in Eqn. \ref{dglobu1} leaves the Lagrangian
invariant.  The $\lambda$ term breaks this down to $Z_4$ as
required. Thus we may identify this term physically with the
(quadrupled) monopole tunnelling events.

In this dual vortex theory, the XY ordered phase is simply
characterized as a dual `paramagnet' where $\langle \psi_{1,2}
\rangle = 0$ and fluctuations of $\psi_{1,2}$ are gapped. On the
other hand, spin paramagnetic phases such as the VBS states correspond to condensates of
the fields $\psi_{1,2}$ which break the dual gauge symmetry. In
particular if both $\psi_1$ and $\psi_2$ condense with equal
amplitude $|\langle\psi_1\rangle |= | \langle\psi_2\rangle |\neq 0$,
then we obtain a paramagnetic phase where the global Ising
symmetry is preserved.  Note the remarkable complementarity
between the description of the phases in this dual theory with
that in terms of the spinon fields of the CP$^1$ representation:
the descriptions map onto one another upon interchanging both $z_{1,2}
\leftrightarrow \psi_{1,2}$ and the role of the XY ordered and
paramagnetic phases. As discussed below, this is a symptom of an
exact duality between the two descriptions that obtains close to
the transition.

\begin{figure}
\centerline{\includegraphics[width=4in]{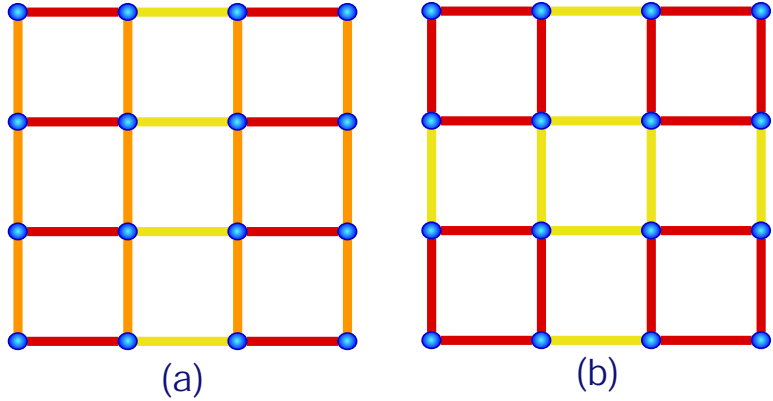}} \caption{Pattern of
symmetry breaking in the two possible VBS states predicted by
(\protect\ref{crtny}). The last term in (\protect\ref{crtny})
leads to a potential, $-\lambda \cos (4 (\theta_1 - \theta_2 ))$,
and the sign of $\lambda$ chooses between the two states above.
The distinct lines represent distinct values of $\langle \vec S_r
\cdot \vec S_{r'} \rangle$ on each link. Note that the state in
(a) is identical to that in Fig~\protect\ref{bond}.} \label{plaq}
\end{figure}

The combination $\psi^*_1\psi_2^{\vphantom*} \equiv |\psi_1\psi_2|
e^{i(\theta_1-\theta_2)}$ actually serves as an order parameter
for the translation symmetry broken VBS ground state. This may be
seen from the analysis of Refs.~\cite{Crtny,sp}.  Alternatively,
we may use the identification, due to Read and
Sachdev\cite{ReSaSuN}, of the skyrmion creation operator with the
order parameter for translation symmetry breaking. Such a
condensate of $\psi_{1,2}$ breaks the global $Z_4$ symmetry of the
action in Eqn. \ref{crtny}. The preferred direction of the angle
$\theta_1 - \theta_2$ depends on the sign of $\lambda$. The two
sets of preferred directions correspond to columnar and plaquette
patterns of translational symmetry breaking (see Fig~\ref{plaq}).

Let us now consider the transition between the XY ordered and VBS
phases described by $\mathcal{L}_{\rm dual}$ at $\lambda = 0$,
{\em i.e} in the absence of instanton events. This model has been
studied \cite{hlm,mv}, and has a remarkable self-dual \cite{mv}
critical point. Here, our arguments demonstrate the self duality
quite simply: recall that we had argued earlier that in the
absence of instantons, the continuum limit of the direct
SU(2)-invariant theory led to $\mathcal{L}_z$ in (\ref{sz}). The
easy plane anisotropy will allow an additional $v |z_1|^2 |z_2|^2$
term in $\mathcal{L}_z$, and then $\mathcal{L}_z$ has exactly the
same form as the dual theory $\mathcal{L}_{\rm dual}$ under
$z_{1,2} \rightarrow \psi_{1,2}$ and $a_\mu \rightarrow A_\mu$!
Note that in both the direct and dual representations, the degrees
of freedom in the Lagrangian are not those associated with the
`physical' boson operator (either $n^+$ or the skyrmion creation
operator). Rather the theory is expressed most simply in terms of
`fractionalized' fields -- namely the spinons or the meron
vortices. In particular, the physical $n^+$ field is a composite
of two spinon fields, and the skyrmion field is likewise a
composite of the two meron fields.

Let us now include monopole events. This is where the existence of
a dual representation pays off, as the non-trivial, non-local
physics of instantons in the direct theory are represented by a
local perturbation in the dual theory: simply set $\lambda \neq 0$
in $\mathcal{L}_{\rm dual}$. The relevance/irrelevance of
monopoles at the self-dual $\lambda = 0$ fixed point is determined
by the scaling dimension, $\Delta$, of the $\left(\psi^*_1
\psi^{\vphantom{*}}_2\right)^4$ operator, {\em i.e} the fourth
power of the creation operator of the physical boson. Provided
$\Delta > d+1 =3$, monopoles will be irrelevant. The $\lambda = 0$
critical fixed point describes an XY ordering transition where the
physical boson field is a composite of the fundamental fields of
the theory.  We therefore expect that correlators of the physical
boson (and its various powers) will decay with a scaling dimension
that is {\em larger} than the corresponding one for the ordinary
XY transition in $D = 2 + 1$ dimensions. Now for the usual XY
fixed point four-fold symmetry breaking perturbations are known to
be {\em irrelevant} \cite{vicari}. This then implies that a small
$\lambda$ will be irrelevant by power counting at the $\lambda =
0$ fixed point of the present model as well.  Note the crucial
role played by the Berry phase term for the monopoles in reaching
this conclusion -- quadrupling the monopoles and thereby
increasing their scaling dimension renders them irrelevant.

Although the $\lambda$ term is irrelevant at the critical fixed
point, it is clearly very important in deciding the fate of either
phase. In particular, in the paramagnetic phase it picks out the
particular pattern of translation symmetry breaking (columnar
versus plaquette) and forces linear confinement of spinons. In
critical phenomena parlance, it may be described as a {\em
dangerously irrelevant} perturbation \cite{csy}.

Thus, in the easy plane case there is the possibility of a direct
second order transition between the N\'{e}el and VBS phases.
Remarkably the critical theory is `deconfined' in the sense that
the spinons emerge as natural degrees of freedom right at the
critical point.  We note that the spinons are {\em confined} in
both phases and do not appear in the excitation spectrum. The
length scale at which this confinement occurs however {\em
diverges} on approaching the critical point.

At a more sophisticated level, the critical fixed point is
characterized by the emergence of an extra global U(1) symmetry
(\ref{dglobu1}) that is not present in the microscopic
Hamiltonian. This is associated with conservation of skyrmion
number and follows from the irrelevance of monopole tunnelling
events only at the critical point.

\section*{Isotropic magnets}

We now provide evidence supporting the possibility that a direct
second transition separates N\'{e}el and VBS states of isotropic ({\em
  i.e.\/} SU(2) invariant) spin-$1/2$ magnets with `deconfinement'
obtaining at the critical point. For this purpose, it is
convenient to work with the direct CP$^1$ representation in terms
of spinon fields. Following Ref.~\cite{SJ}, consider a
generalization to CP$^{N - 1}$ models of an $N$-component complex
field $z_a$ that is coupled to a compact U(1) gauge field with the
same Haldane Berry phases as in the $N = 2$ case of interest. We
argue that both at $N = 1$ and at $N$ large, the model displays a
second-order transition between a `Higgs' phase (this phase is the
analog of the antiferromagnetic N\'{e}el phase for $N=2$) and a
paramagnetic VBS phase with confined spinons which breaks lattice
translation symmetry; moreover, instanton events are irrelevant at
this critical point.

Consider first $N = 1$ where $z \equiv e^{i\phi}$ is simply a
complex number of unit magnitude. This $N = 1$ model displays a
transition between a Higgs and a VBS phase\cite{SJ}. The latter
has a four-fold degenerate ground state due to lattice symmetry
breaking. Simple symmetry arguments suggest a transition modelled
by a $Z_4$ clock model. Since the four fold anisotropy is irrelevant
at the $D = 3$ XY fixed point \cite{vicari}, this is in the $3D$
XY universality class.  Ref.~\cite{SJ} also provided numerical
evidence supporting this expectation.  All of this is also readily
established by the analog of the duality discussed above for
$N=1$. The dual global XY fields simply represent $2\pi$ vortices
in $z$, and the four-fold anisotropy is the quadrupled instanton
event. Once again the quadrupling is due to the Berry phase term.
Thus the irrelevance of the four-fold anisotropy may be
interpreted as the irrelevance of instanton events. The resulting
$3D$ XY universality is simply the dual of the condensation
transition of the $z$ boson coupled to a non-compact U(1) gauge
field \cite{dh}.

Now let us consider $N$ large and begin with the model with {\em
all} monopoles excluded.  This is the non-compact CP$^{N-1}$ model
and has an ordering transition associated with the condensation of
$z$. It is clear that the crucial question is whether the
four-monopole event is relevant/irrelevant at the fixed point of
this non-compact model. The scaling dimension of the $q$-monopole
operator in this model was computed by Murthy and Sachdev\cite{MS}
and their results give a scaling dimension $\propto N$. For large
$N$ this is much larger than $D=3$, and hence the $q=4$ monopoles
are strongly irrelevant for large-$N$\cite{noteN}.

Thus the $N = 1$, $N = \infty$, and easy plane $N = 2$ models all
support the same picture. A direct second order transition between
the spinon-condensed and VBS phases is possible with a
`deconfined' critical point. Right at this point, monopole
tunnelling events become irrelevant, spinon degrees of freedom
emerge as the natural fields of the critical theory, and there is
an extra global conservation law of skyrmion number which is
absent in the microscopic Hamiltonian. This provides strong
evidence that the same scenario obtains for the SU(2) symmetric
model ({\em i.e} at $N = 2$). However, the self-duality found in
the easy plane $N = 2$ model is special and {\em not likely} to
generalize to the isotropic case. In the direct representation,
the critical theory of the isotropic case is the critical point of
$\mathcal{L}_z$ in (\ref{sz}). Remarkably, as claimed earlier, the
complications of compactness and Berry phases (both required by
the microscopics) have cancelled one another, leading ultimately
to a much simpler critical theory~!
Equivalently, the critical theory is that of the $D = 3$ classical
O(3) model with monopoles suppressed.  Direct numerical
simulations of such an O(3) model, and of the non-compact CP$^1$
model, have been performed in Ref.~\cite{mv}, and the results are
consistent with a common critical theory. Earlier work
\cite{dl,km} had examined related O(3) models, and the results for
critical exponents \cite{km} are also consistent with
Ref.~\cite{mv}.

\section*{Physical properties near the `deconfined' critical point}

We now briefly mention the consequences of our theory for the
physical properties near the quantum phase transition between the
N\'{e}el and VBS phases. The presence of a dangerously irrelevant
coupling implies that there are two distinct length scales which
diverge as we approach the critical point from the VBS side: the
spin correlation length $\xi$, and a {\sl longer} length scale
$\xi_{\rm VBS}\sim \xi^{(\Delta-1)/2}$ (where $\Delta
>3$ is the scaling dimension of 4-monopole operator) which
determines the thickness of a domain wall between two VBS states.
Remarkably, on length scales $\xi< L < \xi_{\rm VBS}$, the low
energy excitations of the VBS state are more similar to Goldstone modes
(associated with spontaneous breaking of a continuous symmetry),
than domain walls, despite the discrete broken symmetry of the
phase.
Standard scaling arguments can be used to
predict a variety of physical properties, among which are: ({\em i\/})
the emergence of spinons at the critical point leads to a large
anomalous dimension of the N\'{e}el order, estimated to be $\eta
\approx 0.6$\cite{km,mv}; ({\em ii\/}) the self-duality of the
critical point for the easy plane case implies the remarkable result
that the columnar dimer, plaquette, and staggered XY magnetization all
decay with the same power law at the critical point, and the $\beta$
exponents for the VBS and XY orders are the same.

One intriguing aspect of our theory is the physics of the vortices
of the XY ordered phase (in the easy plane case) close to the
transition.  As discussed extensively above, there are two kinds
of classical meron vortices which tunnel into each other in the
quantum theory. However, the irrelevance of these instanton events
near the transition implies that the tunnelling rate of the Ising
order in the core of merons will diverge on approaching the
transition. This Ising order may be detectable in numerical
studies \cite{sandvik} - this is despite the transition actually
being to a VBS phase with no such staggered order~!

\section*{Discussion}

Our results offer a new perspective on the phases of Mott
insulators in two dimensions: liquid resonating-valence-bond-like
states, with gapless spinon excitations, can appear at isolated
critical points between phases characterized by conventional
confining orders. It appears probable that similar considerations
apply to quantum critical points in doped Mott insulators, between
phases with a variety of spin- and charge-density-wave orders and
$d$-wave superconductivity. If so, the electronic properties in
the quantum critical region of such critical points will be
strongly non-Fermi-liquid like, raising the prospect of
understanding the phenomenology of the cuprate superconductors.

On the theoretical side, our results also illuminate studies of
frustrated quantum antiferromagnets in two dimensions. A theory of
the apparent critical point between the N\'{e}el and VBS phases
observed by Sandvik {\em et al.} \cite{sandvik} is now available,
and precise tests of the values of critical exponents should now
be possible. A variety of other SU(2)-invariant antiferromagnets
have been studied \cite{claire}, and many of them exhibit VBS
phases. It would be interesting to explore the characteristics of
the quantum critical points adjacent to these phases, and test our
prediction of gapless, liquid, resonating-valence-bond-like
behavior.


Our results also caricature interesting
phenomena\cite{piers1,qmsi} found in the vicinity of the onset of
magnetism in the heavy fermion metals. Remarkably the Kondo
coherence that characterizes the non-magnetic heavy Fermi liquid
seems to disappear at the same point that magnetic long range
order sets in. Furthermore strong deviations from Fermi liquid
theory are seen in the vicinity of the quantum critical point. All
of this is in contrast to naive expectations based on the Landau
paradigm for critical phenomena. However this kind of exotic
quantum criticality between two conventional phases is precisely
the physics discussed in the present paper.





\begin{thebibliography}{}
\bibitem{noteN} At large-$N$ the single monopole operator is
  irrelevant, so that even without Berry phases a `deconfined' critical
  fixed point obtains.  However this is special to large-$N$ and does
  not hold at $N = 1$ (or in the easy plane limit at $N = 2$).
\end{thebibliography}
\end{document}